\documentstyle[11pt]{article}
\input epsf

\begin{document}

\begin{titlepage} 
\vspace*{-60pt} 
\begin{flushright} 
SUSSEX-AST 97/3-1 \\ 
gr-qc/9705048\\ 
(March 1997)\\ 
\end{flushright} 
\vspace*{10pt} 
\begin{center} 
\LARGE 
{\bf Can Inflation be Falsified?}\\ 
\vspace{.8cm} 
\normalsize 
\large{John D.~Barrow and Andrew R.~Liddle}\\ 
\normalsize 
\vspace{.6 cm} 
{\em Astronomy Centre, University of Sussex, \\ 
Falmer, Brighton BN1 9QH,~~~U.~K.} 
\end{center} 
 
\vspace{.4 cm} 
\begin{abstract} 
\noindent 
Despite its central role in modern cosmology, doubts are often expressed as 
to whether cosmological inflation is really a falsifiable theory. We 
distinguish two facets of inflation, one as a theory of initial conditions 
for the hot big bang and the other as a model for the origin of structure in 
the Universe. We argue that the latter can readily be excluded by 
observations, and that there are also a number of ways in which the former 
can find itself in conflict with observational data. Both aspects of the 
theory are indeed falsifiable.
\end{abstract} 
 
%%%%%%%%%%%%%%%%%%%%%%%%%%%%%%%%%%%%%%%%%%%%%%%%%%%%%%%%%%%%%%%%%%%%%% 
\end{titlepage} 
%%%%%%%%%%%%%%%%%%%%%%%%%%%%%%%%%%%%%%%%%%%%%%%%%%%%%%%%%%%%%%%%%%%%%% 

%Uncomment to produce double-spaced output 
%\double 

\section{Introduction}

Inflation has become a cornerstone of cosmology --- an enlargement of the
hot big bang theory that is often taken for granted by theorists. But its
venerated position as a paradigm creates nagging doubts about its
predictiveness. Could it ever be ruled out? One of its strongest
advocates, Andrei Linde, has suggested that it cannot be falsified, merely
superseded by a better theory \cite{andreiprinc}. This pessimism derives in
part from fifteen years of exploring the possible outcomes of inflation.
These invariably weaken the original predictions of the theory: for example,
it is now accepted that inflation can lead to a large open Universe, so
undermining the original claim that inflation guarantees a flat Universe and
with it the prospect of falsifying the theory by determining the total
density of the Universe. This elasticity has diminished the faith of the
general astronomical community in inflation, and even led some researchers
to question whether inflationary cosmology is a branch of science at all. In
reality, this gloomy prospect is an overreaction, and the true situation is
much more interesting, as we shall see. But, first we need to distinguish
between two facets of inflation.

In its original incarnation, inflation aimed to solve a number of
cosmological conundrums: to explain why the Universe appears so close to
spatial flatness, why it is so homogeneous and isotropic on large scales,
and why it is not overwhelmed by magnetic monopoles. It is largely in this
first guise that inflation offers little predictive power, although we shall
see that there are several observations which could exclude inflation as an
explanation of even these properties of the Universe. The second facet of
inflation is of far greater cosmological importance and was recognized soon
after its introduction. This is its ability to explain the origin of the
large-scale structure of the Universe. The mechanism it provides is the
stretching of quantum fluctuations to large wavelengths and their subsequent
conversion into classical density perturbations, which seed
structures that can then amplify by gravitational instability. The variety
and precision of the observations that can be used to study the large-scale
structure of the Universe will confront this second facet of the 
inflationary 
universe theory with decisive observational tests.

\section{Inflation as a Theory of Initial Conditions}

Given that inflation was intended to supply a homogeneous, flat,
monopole-free Universe, one might hope that any observation contradicting
this would also exclude inflation. Unfortunately, as far as the observations 
go, we are no
closer to knowing whether the Universe is or is not very close to flatness,
nor have the observational constraints on the admissible monopole density 
been significantly tightened. Instead, the main developments here have
been theoretical, and have served to blunt the sharpness of inflation's 
predictions. There now exist perfectly valid inflation models which
predict a significantly open Universe \cite{open}; ironically, Linde has 
even argued \cite{andreijapan} that an open Universe
is just as strong a piece of evidence in favour of inflation as
is an almost flat Universe, because inflation is the only known way of
creating an open Universe that is also homogeneous. Even the monopole 
problem is no longer clear cut.
There exist perfectly
valid inflation models (e.g.~Ref.~\cite{yok}) which can predict any monopole
density from zero up to the present observational upper limits (and 
above). Discovery of a monopole density somewhere {\it below} current 
observational limits would also support inflation, because low monopole 
number densities imply a violation of causality in the 
absence of inflation (a minimum number density well above the present 
observational limits is required by the Kibble mechanism and the known 
properties of monopoles \cite{Pres}). It is clear that the observation of a 
hyperbolic geometry, or a non-zero monopole density, will not falsify 
inflation.

String theorists expect that there might exist long-lived Planck mass relics
from the Planck or string scale. Clearly, unless their production were
extraordinarily suppressed below thermal expectations, these relics would
soon come to dominate the density of the Universe and create a similar
problem to that posed by monopoles. If such superheavy particles were found
they would tell us that inflation had not occurred to dilute their
abundance. Likewise, any specific quantum cosmological signature of boundary
conditions at the `initial' quantum state of the Universe would be erased
and overwritten by inflation, but could persist to the present if inflation
did not occur. Thus inflation, whilst good news for cosmologists, is
extremely bad news for the study of Planck scale physics: any information of
cosmological events at the Planck scale is degraded to unobservably low
levels by the occurrence of inflation. If you seek to provide an explanation
of the Universe's structure independent of initial conditions, you cannot
expect to learn anything of those initial conditions from observations of
its structure. Testing any theory of Planck scale quantum gravity is always 
going to be harder than testing inflation.

The most decisive observational evidence against inflation would be provided
by evidence that the Universe possesses large-scale rotation. Any rotation
existing prior to inflation is reduced to unobservably low levels by
inflation, as the various cosmic no-hair theorems demonstrate. The same is
true for pre-inflationary density irregularities or gravitational-wave
distortions, but whereas density perturbations (scalar modes) and
gravitational waves (tensor modes) can be generated during inflation to seed
the structure of the Universe in the future, rotational perturbations
(vector modes) cannot. The scalar nature of the source of inflation is
needed to ensure that slow evolution (slower than the Hubble rate) is
possible. Moreover, the motion of scalar fields is necessarily irrotational. 
Only upper limits on cosmic rotation exist at present~\cite{rot}, but if
characteristic signatures of large-scale vorticity were to be found in 
microwave background maps then this would be incompatible with inflation. 
Vorticity
cannot exist without shear distortion and observable levels of shear
anisotropy would also be embarrassing for inflationary theories.
Interestingly, it has been shown that physically realistic initial
conditions of the sort used to study inflationary universe models predict a
maximum level of microwave background temperature anisotropy that is very
close to the observed level on large angular scales \cite{Bar}. If the 
signature of
these homogeneous anisotropies were detected it would rule out
the occurrence of inflation in the past. When the expansion is isotropic,
these anisotropies, which are contributed by very long-wavelength
gravitational waves, will be absent and there should only exist a thermal
graviton background as a relic of the Planck scale. This would not survive
inflation, but if inflation did not occur it would still lie many orders of 
magnitude below the sensitivity of
planned gravitational-wave detectors.

The discovery of observational evidence for any non-trivial topology of the 
Universe at or below the horizon would indicate that inflation has not 
occurred in the past, as the periodicity scale would be pushed beyond the 
horizon if inflation solved the homogeneity problem. Current microwave 
observations require any periodic topology scale to be at least of order the 
horizon scale \cite{top}. The existence of complex (`fractal') topology of 
space-time foam on sub-Planck scales ($\ll 10^{33}$ cm) would also probably 
make it difficult for inflation to supply large-scale smoothness. 

\section{Inflation as a Theory of the Origin of Structure}

In the 1980s, it became the accepted folklore that inflation predicts a
scale-invariant spectrum of density perturbations, and inflation might 
appear 
to have been undermined by the discovery that it can predict a range of
different spectra. In fact, the original prediction was of only approximate 
scale invariance (Bardeen et al.'s paper \cite{BST} was entitled {\it
Spontaneous Creation of Almost Scale-Free Density Perturbations in an
Inflationary Universe}), commensurate with the quality of the data at that
time. The remarkable improvement in the quality of the astronomical data has
made such generalized predictions inadequate. A higher order of predicted
detail is now required. Unfortunately, at this higher level of precision,
the realization that inflation makes a broader set of predictions has led to
an impression that inflation can in fact predict anything at all. This is
far from the truth.

As Albrecht has emphasised \cite{andy}, the fact that
inflationary perturbations are laid down in the distant past and evolve
passively, through linear theory, up to the present endows them with
distinctive generic features. Since there is invariably a growing mode which
dominates long before a given scale enters the horizon, those scales short
enough to exhibit oscillatory behaviour when they enter the horizon will
display oscillations with fixed temporal phases. This leads ultimately to
the familiar oscillatory character of the radiation angular power spectrum,
such as that in Figure~1. The oscillatory prediction has been pushed
strongly as a test of inflation in Ref.~\cite{HW}. If the power spectrum
observed by the MAP or Planck Surveyor satellites fails to display these 
oscillatory
features (it seems likely, for instance, that the cosmic string theory
predicts their absence \cite{imps}) then inflation will be in conflict with
observation.

\begin{figure}[tbp]
\centering 
\leavevmode\epsfysize=10cm \epsfbox{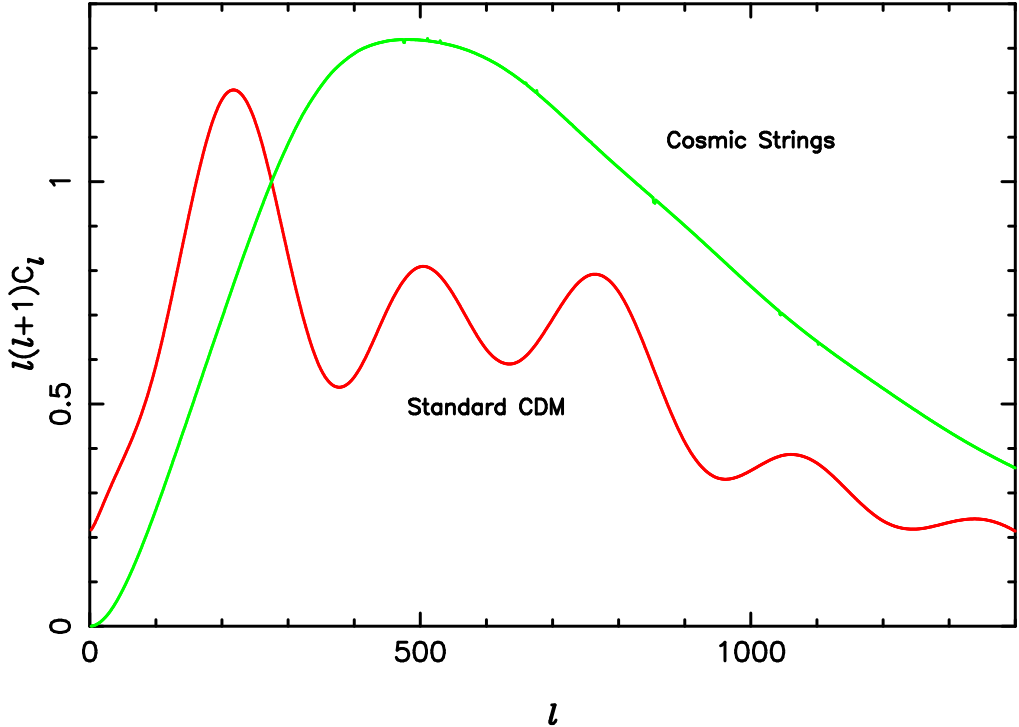}\\
\caption[fig1]{The radiation angular power spectrum for standard CDM, based
on inflation, and that estimated for a cosmic string scenario \cite{imps}.
The relative normalization is arbitrary; $C_\ell$ and $\ell$ have their
usual meaning.}
\label{fig1}
\end{figure}

Ultimately, the angular power spectrum arises from linear theory, so one
might question whether or not it is possible to choose a very complicated
initial power spectrum, so as to mimic precisely the observed properties, at
the percent or so level, of any observations that might be made in the
future. Leaving aside all aesthetic prejudices about such theoretical
over-engineering of the initial conditions, it is actually impossible in the
case of inflation. One cannot insert arbitrarily sharp features into the
power spectrum if inflation is occurring. As an example, Figure~1 shows a
standard cold dark matter (CDM) spectrum, generated by a scale-invariant
inflationary spectrum, alongside a cosmic string spectrum from 
Ref.~\cite{imps}. In order to produce the latter from an inflationary model, 
the input
power spectrum from inflation would have to look like that shown in Figure~2 
(which covers $\ell $ from 50 upwards, the region where the string spectrum 
is valid). The
lower panel shows that the required spectral index would have to fluctuate
dramatically between approximately $-4$ and $6$. Such values are
incompatible with inflation taking place: it is impossible to tinker with
the initial power spectrum sufficiently to reproduce observations of this
sort. They would be fatal to inflation. One cannot transform a standard CDM
spectrum into a cosmic string spectrum. Moving to a low-density Universe
will match the position of the peak better, but the need to erase the
oscillations by introducing extra out-of-phase oscillations in the initial
power spectrum to cancel them still remains.

\begin{figure}[tbp]
\centering 
\leavevmode\epsfysize=10cm \epsfbox{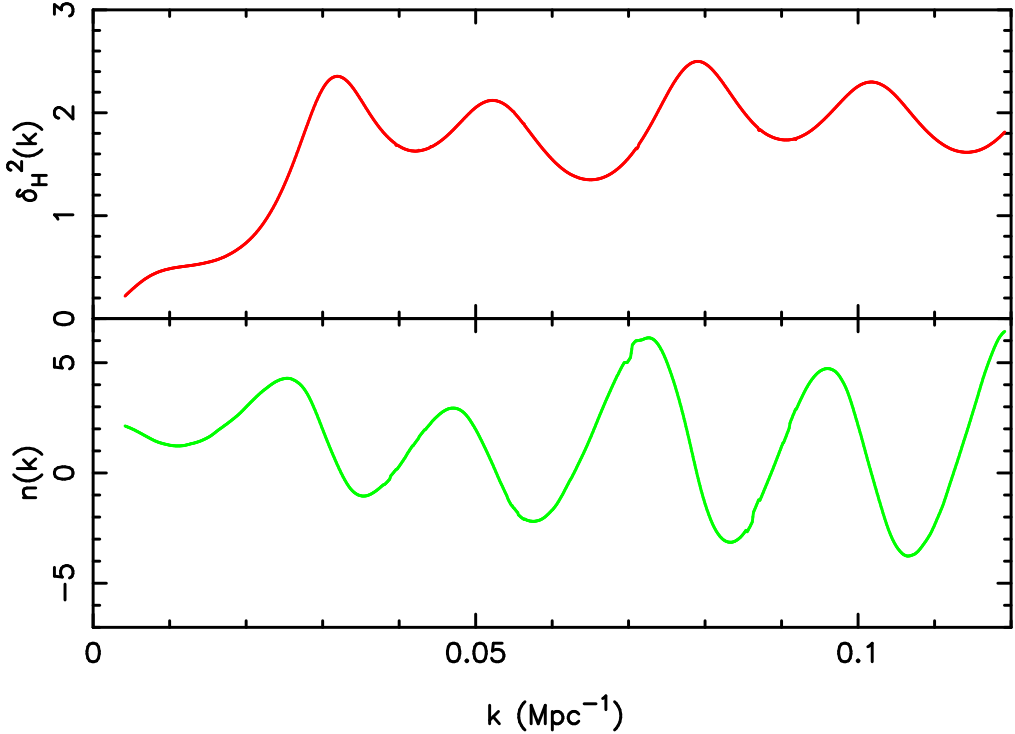}\\
\caption[fig2]{The required power spectrum for the inflation model to mimic
the string model (arbitrarily normalized). The upper panel shows the power 
spectrum, defined as in Ref.~\cite{PhysRep}, and the lower one the spectral 
index (the minor glitches are from rounding errors in the uncertain string 
prediction).}
\label{fig2}
\end{figure}

Any proof of the existence of large-scale perturbations on scales well
above the Hubble radius has already been cited as a strong indicator of
inflation \cite{iatuw} --- one can demonstrate that their existence (not yet
required by existing data) would imply either that perturbations were
generated acausally, or that a period of inflation occurred. Note that this
is true regardless of whether or not inflation produces the perturbations.
This would be supporting evidence for inflation, but it can also be turned 
on its head and used to rule out inflation as the explanation of structure 
formation.
While we see that inflation must generate perturbations on scales close to 
the
present horizon (we exist!), inflation cannot know where {\em our} present
horizon will be. Therefore, a generic prediction of inflation is that
there {\it must} exist perturbations on scales larger than the Hubble
length. This can be tested at decoupling using the microwave background, 
and conceivably at earlier times using nucleosynthesis. If 
these can be shown not to exist, inflation as a model of structure formation 
will be falsified.

The details of the perturbations may also inveigh against inflation. In
single-field inflation models, there is a relation between the scalar and
tensor modes that inflation produces \cite{cons}. Confirmation of this 
relation would be strong evidence in favour of single-field inflation. 
Unfortunately this relation does not survive as an equality in the case 
where there are many fields driving inflation, but it does survive as an 
inequality \cite{2field} which could certainly be violated by actual 
observations. The existence of vortical or magnetic field perturbations on 
very large scales would also be impossible to accommodate in present 
inflationary models, which are unable to support perturbations of a vector 
nature, though it is possible that models could be constructed giving 
magnetic perturbations by introducing new couplings \cite{mag}.

The general statistics of inhomogeneous perturbations do not seem so useful 
as a test of inflation. In the simplest models, one has the powerful 
limitation that the perturbations are gaussian and adiabatic, both eminently 
testable,
but each of these possibilities can be violated in more complex models.
However, the only `reasonable' non-gaussian models presently
in existence are simple transformations of a gaussian
(typically a chi-squared distribution). Observational
evidence for an initial statistical distribution of fluctuations that is
significantly more complex (such as that expected from topological defects)
could still eliminate inflation as an explanation for structure formation.

\section{Verdict}

In summary, we believe that as a model of initial conditions, inflation is a 
very adaptable theory, but
there remain several ways in which it might be ruled out. It remains to be
seen whether any of these tests become decisive. But as a model of structure
formation, inflation lives much more dangerously. Future observations offer
the prospect of a critical test. Whether inflation created the large-scale
structure of the Universe is at present not proven, but will eventually be
decided, one way or the other, beyond all reasonable doubt.

%%%%%%%%%%%%%%%%%%%%%%%%%%%%%%%%%%%%%%%%%%%%%%%%%%%%%%%%%%%%%%%%%%%%%% 

\section*{Acknowledgments}

JDB is supported by the PPARC and ARL by the Royal Society.

%%%%%%%%%%%%%%%%%%%%%%%%%%%%%%%%%%%%%%%%%%%%%%%%%%%%%%%%%%%%%%%%%%%%%% 
\frenchspacing 
%%%%%%%%%%%%%%%%%%%%%%%%%%%%%%%%%%%%%%%%%%%%%%%%%%%%%%%%%%%%%%%%%%%%%% 

%%%%%%%%%%%%%%%%%%%%%%%%%%%%%%%%%%%%%%%%%%%%%%%%%%%%%%%%%%%%%%%%%%%%%% 

\end{document}